\documentclass[12pt,onecolumn,journal]{IEEEtran}
\usepackage{amsmath,amsfonts, amssymb, mathrsfs}
\usepackage[dvips]{graphicx}
\usepackage[dvips]{color}

\newcommand{\ol}[1]{\overline{{#1}}}
\newcommand{\argmin}{\operatornamewithlimits{arg\ min}}

\newtheorem{remark}{Remark}[section]
\newtheorem{theorem}{Theorem}[section]
\newtheorem{proposition}{Proposition}[section] 

\title{On feedback in Gaussian multi-hop networks\vspace{-3mm}}
\author{\IEEEauthorblockN{Bobbie Chern, Farzan Farnia, Ayfer \"{O}zg\"{u}r} \\
\IEEEauthorblockA{Stanford University\\
\{bgchern, farnia, aozgur\}@stanford.edu}\vspace{-6mm}
\thanks{This work was supported in part by National Defense Science \&
Engineering Graduate Fellowship (NDSEG) and NSF CAREER award 1254786.  This work was presented in part in the Information Theory Workshop in Seville in 2013 and the Information Theory and Applications Workshop in San Diego in 2014.}}

\begin{document}
\maketitle
\begin{abstract}
The study of feedback has been mostly limited to single-hop communication
settings. In this paper, we consider Gaussian networks where sources and
destinations can communicate with the help of intermediate relays over multiple
hops. We assume that links in the network can be bidirected providing
opportunities for feedback. We ask the following question: can the information
transfer in both directions of a link be critical to maximizing the end-to-end
communication rates in the network? Equivalently, could one of the directions
in each bidirected link (and more generally at least one of the links forming a
cycle) be shut down and the capacity of the network still be approximately
maintained? We show that in any arbitrary Gaussian network with bidirected
edges and cycles and unicast traffic, we can always identify a directed acyclic
subnetwork that approximately maintains the capacity of the original network.
For Gaussian networks with multiple-access and broadcast traffic, an acyclic
subnetwork is sufficient to achieve every rate point in the capacity region of
the original network, however, there may not be a single acyclic subnetwork that
maintains the whole capacity region. For networks with multicast and multiple
unicast traffic, on the other hand, bidirected information flow across certain links
can be critically needed to maximize the end-to-end capacity region. These
results can be regarded as generalizations of the conclusions regarding the
usefulness of feedback in various single-hop Gaussian settings
and can provide opportunities for simplifying operation in Gaussian multi-hop
networks.  
\end{abstract}

\section{Introduction}
Feedback has been studied extensively for single-hop communication channels.
While feedback cannot increase the capacity of the discrete memoryless
point-to-point channel \cite{Shan56}, it is well understood that it can
increase the capacity of the Gaussian multiple access (MAC), broadcast and
relay channels,  but only through a power gain \cite{EGC79,Ozarow84,Thomas87}. More
recently, it has been shown in \cite{SuhTse11} that feedback can provide
degrees of freedom gain in the Gaussian interference channel, which translates
to an unbounded gain in capacity when SNR increases. In the recent years, there
has been significant interest in larger networks where communication between
nodes is established in multiple hops
\cite{OLT07,AvDigTse09,KoetterEffrosMedard}. However, the study of the
usefulness of feedback has been mostly limited to the above single-hop settings
of a few nodes. 

In this paper, we aim to understand the role of feedback in general Gaussian
networks. We consider a Gaussian network where sources communicate to
destinations in multiple-hops with the help of intermediate relay nodes. In
wireless, if a given node can send information to another node, typically it
can also receive information from that node, thus communication links between
pairs of nodes are often bidirectional. Therefore, inherently there are a lot
of opportunities for ``feeding back'' information in wireless networks, though
the nature of these feedback links is significantly different from the
idealized feedback models considered in the single-hop settings. First,
transmissions, and therefore also feedback, may not be isolated but subject to
broadcast and superposition. Second, while in single hop networks the links
originating from destinations and/or arriving at source nodes can be clearly
identified as feedback, in multihop networks there can be ``feedback'' between
any pair of nodes. Bidirected links and cycles in the network can be used to
feedback information, however it is not a priori possible to designate links as
communication links and feedback. Therefore, in these new multi-hop settings it
is not totally clear how to think about feedback and how to study its
usefulness.
\begin{figure}[t]
	\begin{center}
		\includegraphics[scale=1.0]{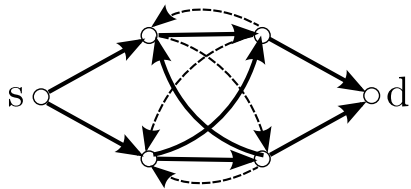}
	\end{center}
  \caption{Bidirected network with the directed acyclic subnetwork carrying approximately its capacity highlighted.}
	\label{fig:twolayers1}
\end{figure}

In this paper, we adopt the following approach. We consider a general Gaussian
relay network with arbitrary topology and channel gains, possibly with
bidirected links and cycles, where some links can be subject to broadcast and
superposition and some can be isolated in a completely arbitrary fashion. We
ask the following question: can the information transfer in both directions of
any link in the network be critical to maximizing the end-to-end communication
rate? Equivalently, could one of the directions in each bidirected link (and
more generally at least one of the links forming a cycle) be shut down and the
capacity of the network still be approximately maintained?

We show that when there is only a single source-destination pair in the network
(unicast traffic), we can always identify a directed acyclic subnetwork that
approximately preserves the capacity of the original network. More precisely,
if any of the links that do not belong to this subgraph could be disabled, the
capacity of the resultant network would still remain within a bounded gap to the
capacity of the original network. See Figure~\ref{fig:twolayers1}. The main
technical step is to show that in every Gaussian relay network, there exists a
directed acyclic subnetwork for which the information theoretic cutset upper
bound evaluated under i.i.d. input distributions is exactly the same as that
for the original network. 

Conceptually, identifying the directed acyclic ``skeleton'' subnetwork that approximately carries the capacity of the network can be used to classify links as information carriers (critical to information transfer) and feedback links (of limited contribution to capacity). It also allows one to associate a direction with the information flow in an undirected wireless network.  From a practical perspective, this result provides possibilities for simplifying network operation (in terms of delay and
complexity) by identifying feedback links that can be potentially shut down without
significantly impacting capacity. The simplification is immediate in networks with isolated links, such as graphical networks, which form a special case of the model we consider in this paper. For wireless networks, shutting down individual links may be nontrivial since these links may represent interference or overheard transmissions over other links. In Section~\ref{sec:discuss}, we provide examples which illustrate possibilities for simplifying network operation even in the wireless case. This simplification aspect of our result is similar in spirit
to \cite{simplification} and \cite{michelle}, where  \cite{simplification}
seeks a high-capacity small core in a wireless relay network that carries a
good fraction of the overall capacity and \cite{michelle} investigates the
impact of removing a single edge on the capacity region of a graphical (wired) network.

After discussing the unicast case, we extend our result to more general traffic models. We show that for multiple-access (multiple sources communicating to the same destination node)
and broadcast (a single source node communicating independent information to
multiple destinations) traffic, each rate point in the capacity region of the
original network can be approximately achieved by using an acyclic directed
subnetwork. However, a single acyclic directed subnetwork that allows to
approximately achieve all the rate points in the original capacity region may
not exist. For multicast (a single source node communicates the same
information to multiple destinations) and multiple unicast (multiple
source-destination pairs communicating independent information with each other)
traffic, we provide examples which illustrate that bidirected communication
over certain links can be critical to achieving capacity. These results provide
a generalization of the conclusions for the single-hop case, where it is known
that the capacity gain from feedback is bounded (or absent) for point-to-point,
multiple-access and broadcast Gaussian channels and can be unbounded in the case of
interference channels.

We state the main results of our paper in Section~\ref{sec:mainresult},
prove them in Sections~\ref{sec:unicast} and \ref{sec:MACandBC} and discuss their
implications in more detail in Section~\ref{sec:discuss}.

\section{Model}\label{sec:model}
We consider a bidirected Gaussian relay network $G$ consisting of a set of
nodes $V$ and communication links $E$.  We let $|V|$ denote the total number of
nodes.   All nodes in the network are able to send and receive, thus, for each
pair of nodes $u,v \in V$ we can potentially have links $(u,v) \in E$ and
$(v,u) \in E$ with arbitrary channel gains. We assume the links with non-zero
channel gains are represented with directed edges as in
Figure~\ref{fig:twolayers1} giving rise to a directed graph with potentially
bidirected edges and cycles. We assume nodes can have multiple transmit and
receive antennas. Let $X_v\in \mathbb{C}^{M_v}$ denote the signal transmitted
by node $v \in V$ with $M_v$ transmit antennas.  Similarly, let $Y_v\in
\mathbb{C}^{N_v}$ denote the signal received by node $v \in V$ with $N_v$
receive antennas. We have 
\begin{align*}
Y_v = \sum_{u \in V}H_{vu}X_u + Z_v,
\end{align*}
where $H_{vu}$ denotes the channel matrix from node $u$ to node $v$. This
multiple-input multiple-output channel model can also be used to incorporate
networks where different channels operate on different frequencies as well as
networks with isolated links.
\footnote{Indeed, the conclusions of the paper also hold for wired networks (in
this case with no gap) and a mixture of wireless and wired networks.}
The noise $Z_v$ are independent and circularly symmetric Gaussian random
vectors $\mathcal{N}(0,I)$.  All nodes are subject to an average power
constraint $P$. Note that the equal power constraint assumption is without loss
of generality as the channel coefficients are arbitrary. 

We consider the following traffic scenarios over this network:
\begin{itemize}
\item \text{Unicast:} Source node $s\in V$ wants to communicate to  destination
  node $d\in V$. The capacity of the network $G$, denoted by $C(G)$, is the
  largest rate at which $s$ can reliably communicate to $d$.
\item \text{Multiple-Access:} Source nodes $s_1, s_2,\dots, s_n\in V$ want to
  communicate independent messages to a destination node $d\in V$. The capacity
  region $C(G)$ is the closure of jointly achievable rate pairs $R_1, \dots,
  R_n$ where $R_i$ is the reliable communication rate from $s_i$ to $d$.
\item \text{Broadcast:} Source node $s\in V$ wants to communicate independent
  messages to destination nodes $d_1, \dots, d_n\in V$. The capacity region
  $C(G)$ is the closure of jointly achievable rate pairs $R_1, \dots, R_n$
  where $R_i$ is the reliable communication rate from $s$ to $d_i$.
\item \text{Multicast:} Source node $s\in V$ wants to communicate the same
  message to destination nodes $d_1, \dots, d_n\in V$. The capacity $C(G)$ is
  the largest rate $R$ at which the message can be simultaneously communicated
  to all destinations.
\item \text{Multiple-Unicast:} Source node $s_i\in V$ wants to communicate an independent message to its destination node $d_i\in V$ for $i=1,\dots, n$. The capacity region
  $C(G)$  is the closure of jointly achievable rate pairs $R_1, \dots, R_n$
  where $R_i$ is the reliable communication rate from $s_i$ to $d_i$. 
\end{itemize}

Note that we slightly abuse notation here by  using $C(G)$ to refer to a single number in the case of unicast and multicast traffic and a region in the case of multiple-access, broadcast and multiple-unicast traffic. The same is true for $C_{i.i.d.}(G)$ we define in the next section. The usage should be clear from the context.

\section{Main Results}\label{sec:mainresult}
For an arbitrary bidirected Gaussian relay network $G$ with a set of nodes $V$
and communication links $E$, we define a directed acyclic subnetwork
$\tilde{G}$ to be one which consists of the same set of nodes $V$ and a subset
of the communication links $\tilde{E} \subseteq E$. For the Gaussian relay
network, this corresponds to setting the channel coefficients corresponding to
the edges in $E\setminus \tilde{E}$  to zero. A directed acyclic subnetwork
satisfies the property that for any pair of nodes $u,v \in V$, if $(u,v) \in
\tilde{E}$ then $(v,u) \not \in \tilde{E}$.  In other words, if there is a link
in one direction between any two nodes, there cannot be a link in the opposite
direction.  Moreover, it contains no cycles. That is, for every set of nodes
$v_1, \ldots, v_N \in V$, at least one of the edges $(v_1, v_2),\ldots, (v_{k},
v_{k+1}), \ldots, (v_N, v_1) \not \in \tilde{E}$ for any value of
$N$.\looseness=-1

The main conclusions of this paper are summarized in the following three theorems.
\smallbreak
\begin{theorem}\label{thm:mainthm1}
Let $C(G)$ be the capacity of a Gaussian network $G$ with unicast traffic.  We
can identify a directed acyclic subnetwork $\tilde{G}$ of $G$ whose capacity
$C(\tilde{G})$ in bits/s/Hz is bounded by
$$
C(G) - g_1 \leq C(\tilde{G}) \leq C(G) + g_1
$$ 
where $g_1$ is a constant independent of the channel gains and SNRs and can be upper bounded by $3.3 M$ where $M=\sum_{v\in V}M_v+ N_v$ is the total number of antennas in the network.
\end{theorem}

\smallbreak
The fact that the gap between the capacity of $\tilde{G}$ and that of the original network $G$ can be bounded independent of the channel gains and SNRs  implies that the gain due to using the additional links in $G$ remains bounded as SNR grows. The core of our argument for proving this theorem is summarized in the following proposition, which indeed only involves the information-theoretic cutset upper bound on the capacity of the network evaluated under i.i.d. input distributions, denoted by $C_{i.i.d.}$. We show that $C_{i.i.d.}$ for $G$ and $\tilde{G}$ are equal to each other without any gap. One way to interpret $C_{i.i.d.}$ is as an upper bound on the capacity of the network when no beamforming strategies are allowed. In turn, the fact that $C_{i.i.d.}$ is the same for $G$ and $\tilde{G}$ can be interpreted as follows: when beamforming strategies are not allowed the additional feedback links in $G$ can not provide any capacity gain. Equivalently the feedback links in $G$ can only provide additional rate gain through beamforming. However, note that this interpretation is not totally exact since $C_{i.i.d.}$ is only an upper bound on the capacity of the network with no beamforming strategies and not the exact capacity which remains unknown. 

\smallbreak
\begin{proposition}\label{prop0}
Consider a Gaussian network $G$ with unicast traffic. Let
\begin{equation}\label{eq:cutfunction0}
C_{i.i.d.}(G)=\min_{S} f(G;S),
\end{equation}
where $S\subseteq V: s\in S, d\notin S$ is a source-destination cut of the
network and $f(G;S)$ for all $S\subseteq V$ is defined as
\begin{equation}\label{eq:cutfunction}
f(G;S)=I(X_{S};Y_{S^c}\lvert X_{S^c}),
\end{equation}
where $X_v,\,v\in V$ are i.i.d. $\mathcal{CN}(0,(P/M_v)I)$. In other words, $C_{i.i.d.}(G)$ is the information theoretic cutset upper bound on the capacity of the network evaluated under an i.i.d. Gaussian input distribution. Then in every bidirected network $G$ with $C_{i.i.d.}(G)$, we can identify a directed acyclic
subnetwork $\tilde{G}$ with $C_{i.i.d.}(\tilde{G})= C_{i.i.d.}(G)$.
\end{proposition} 

\smallbreak
The proof of Theorem~\ref{thm:mainthm1} follows by combining this proposition
with the existing results in the literature which show that the capacity $C(G)$
of any Gaussian relay network with unicast traffic is within a bounded gap to
$C_{i.i.d.}(G)$ \cite{AvDigTse09,OD10,LKEC11}. We recall the following result from
\cite{LKEC11}:
\begin{theorem}[Theorem 4, \cite{LKEC11}]\footnote{The result in \cite{LKEC11} is stronger than what is stated here as it shows that $C(G)\geq \bar{C}(G)- g_2$ where $\bar{C}$ is the actual information-theoretic cutset upper bound on the capacity of the network. We use the weaker form $C(G)\geq C_{i.i.d.}(G)- g_2$ here as we need a lower and an upper bound on $C(G)$ in terms $C_{i.i.d.}(G)$ in order to connect Theorem~\ref{thm:mainthm1} and Proposition~\ref{prop0}. Also, the result in \cite{LKEC11} is for Gaussian channels with real input and outputs. The gap stated here is for complex channels  which is twice the gap for the real case.} 
In any Gaussian relay network $G$ with unicast traffic, we can achieve all rates
\begin{align}
R\leq C_{i.i.d.}(G)- g_2\label{eq:cutset}
\end{align}
where $g_2\leq 1.3M$. Equivalently, $C(G)\geq C_{i.i.d.}(G)- g_2$.
\end{theorem}

It has been also shown in Lemma 6.6. of \cite{AvDigTse09} that the restriction to i.i.d. Gaussian input distributions is within $g_3=2M$ bits/s/Hz of the actual information-theoretic cut-set upper bound $ \bar{C}(G)$, i.e. for any Gaussian network $G$
\begin{equation}\label{eq:approxcap}
C(G)\leq \bar{C}(G)\leq C_{i.i.d.}(G)+g_3.
\end{equation} 
This shows that within a total gap of $g_2+g_3$, the capacity of the network is approximately given by $C_{i.i.d.}(G)$. More precisely,
\begin{equation}\label{eq:approxcap1}
C_{i.i.d.}(G)-g_2\leq C(G)\leq C_{i.i.d.}(G)+g_3.
\end{equation}
The proof of Theorem~\ref{thm:mainthm1} follows immediately by combining \eqref{eq:approxcap1} with Proposition~\ref{prop0} where $g_1=g_2+g_3$.

\bigbreak
The following theorems state analogous results for the multiple access and broadcast case. The proofs of these theorems follow a similar structure to the unicast case.

\begin{theorem}\label{thm:mainthm2}
Let $C(G)$ be the capacity region of a Gaussian network $G$ with multiple access traffic.
If $(R_1,R_2, \dots ,R_n)\in C(G)$, then there exists an acyclic subnetwork $\tilde{G}$ of $G$ 
such that $$(R_1-g_1,R_2-g_1, \dots ,R_n-g_1)\in C(\tilde{G}),$$ where 
$g_1$ is a constant independent of the channel gains and SNRs. 
$g_1$ can be upper bounded by $3.3 M$.
\end{theorem}

\smallbreak
\begin{theorem}\label{thm:mainthm3}
Let $C(G)$ be the capacity region of a Gaussian network $G$ with broadcast traffic.
If $(R_1,R_2, \dots ,R_n)\in C(G)$, then there exists an acyclic subnetwork $\tilde{G}$ of $G$
such that $$(R_1-g_4,R_2-g_4, \dots ,R_n-g_4)\in C(\tilde{G}),$$ 
where $g_4$ is a constant independent of the channel gains and SNRs. 
$g_4=O(M\log M)$ where $M$ again is the total number of antennas in the network. 
\end{theorem}

Analogous to the unicast case, the proofs of these theorems are based on the following propositions which only involve the information-theoretic cutset upper bound on the capacity region of the network when evaluated under i.i.d. input distributions.

\smallbreak
\begin{proposition}
\label{prop1}
Consider a Gaussian network $G$ with multiple access traffic. Let
$C_{i.i.d.}(G)$ be the set of rate tuples $(R_1,R_2, \dots ,R_n)$ such that 
\begin{equation}\label{eq:cutfunction1}
\sum_{s_i\in S} R_i \leq f(G;S),
\end{equation}
$\forall S\subseteq V: d\notin S$ and $f(G;S)$ is defined in \eqref{eq:cutfunction}. Then for each
$(R_1,R_2, \dots ,R_n) \in C_{i.i.d.}(G)$, we can identify a directed acyclic
subnetwork $\tilde{G}$ of $G$ such that $(R_1,R_2, \dots ,R_n) \in
C_{i.i.d.}(\tilde{G})$ (where $C_{i.i.d.}(\tilde{G})$ is defined analogously
according to \eqref{eq:cutfunction1} for $\tilde{G}$).
\end{proposition}

\smallbreak
\begin{proposition}
\label{prop2}
Consider a Gaussian network $G$ with broadcast traffic. Let $C_{i.i.d.}(G)$ be the set of rate tuples $(R_1,R_2, \dots ,R_n)$ such that 
\begin{equation}\label{eq:cutfunction2}
\sum_{d_i\notin S} R_i\leq  f(G;S),
\end{equation}
 $\forall S\subseteq V: s \in S$ and $f(G;S)$ is given in \eqref{eq:cutfunction}.
Then for each $(R_1,R_2, \dots ,R_n) \in C_{i.i.d.}(G)$,
we can identify a directed acyclic subnetwork $\tilde{G}$ of $G$ such that
$(R_1,R_2, \dots ,R_n) \in C_{i.i.d.}(\tilde{G})$ (where
$C_{i.i.d.}(\tilde{G})$ is defined analogously according to \eqref{eq:cutfunction2} for $\tilde{G}$).
\end{proposition}

\smallbreak
\begin{remark} Note that Propositions~\ref{prop1} and \ref{prop2} do not imply that $C_{i.i.d.}(G)\subseteq C_{i.i.d.}(\tilde{G})$, since the subgraphs $\tilde{G}$ which we identify here may not be the same for different rate points $(R_1,R_2, \dots ,R_n) \in C_{i.i.d.}(G)$. In other words, in both cases there may not be a single subnetwork $\tilde{G}$ which achieves all the rate points $(R_1,R_2, \dots ,R_n) \in C_{i.i.d.}(G)$. In Section~\ref{sec:MACandBC}, we provide examples which illustrate this point.
\end{remark}

\smallbreak
The proofs of Theorems~\ref{thm:mainthm2} and \ref{thm:mainthm3} similarly follow by
combining Propositions~\ref{prop1} and \ref{prop2} respectively with the
existing results in the literature which show that the capacity region of a
Gaussian network $G$ with multiple-access \cite{LKEC11} or broadcast
\cite{broadcast} traffic is within a bounded gap to $C_{i.i.d.}(G)$. We restate Theorem 4 of \cite{LKEC11} now in its more general form which holds for multiple-access traffic and also recall the main result of \cite{broadcast} for broadcast traffic. 
\smallbreak
\begin{theorem}[Theorem 4, \cite{LKEC11}]\label{thm:MAC}
Consider any Gaussian network with multiple-access traffic and let $(R_1,R_2, \dots
,R_n)\in C_{i.i.d.}(G)$. Then $(R_1-g_2,R_2-g_2, \dots ,R_n-g_2)\in C(G)$ where
$g_2\leq 1.3 M$.
\end{theorem}

\smallbreak
\begin{theorem}[Theorem 1, \cite{broadcast}]\label{thm:BC}
Consider any Gaussian network with broadcast traffic and let\\
 $(R_1,R_2, \dots ,R_n)\in C_{i.i.d.}(G)$. Then $(R_1-g_5,R_2-g_5, \dots ,R_n-g_5)\in C(G)$ where $g_5=O(M\log M)$.
\end{theorem}

\smallbreak
For broadcast and multiple-access traffic, the fact that the restriction to i.i.d. Gaussian
input distributions is within $g_3=2M$ bits/s/Hz of the actual
information-theoretic cut-set upper bound $ \bar{C}(G)$ in Lemma 6.6 of \cite{AvDigTse09} implies that
\begin{equation}\label{eq:approxcap2}
C(G)\subseteq \bar{C}(G)\subseteq C_{i.i.d.}(G)+g_3,
\end{equation}
which implies that for any $(R_1,R_2, \dots ,R_n)\in C(G)$,
$(R_1-g_3,R_2-g_3, \dots ,R_n-g_3)\in C_{i.i.d.}(G)$. Together with the
results in the last two theorems, this implies that within a gap independent of the channel gains and
SNRs, the capacity region of a Gaussian network with multiple access and
broadcast traffic is approximately given by $C_{i.i.d.}(G)$. 

The proof of Theorems~\ref{thm:mainthm2} and \ref{thm:mainthm3} for the
multiple access and broadcast traffic scenarios follow immediately by combining
Theorems~\ref{thm:MAC} and \ref{thm:BC} and Eq.~\eqref{eq:approxcap2} with
Propositions~\ref{prop1} and \ref{prop2}. Let $(R_1,R_2, \dots ,R_n)$ be in the
capacity region of the original Gaussian network $G$ with multiple access
traffic.  \eqref{eq:approxcap2} implies that $(R_1-g_3,R_2-g_3, \dots
,R_n-g_3)\in C_{i.i.d.}(G)$. In turn, Proposition~\ref{prop1} implies that
there exists a directed acyclic subnetwork for which $(R_1-g_3,R_2-g_3, \dots
,R_n-g_3)\in C_{i.i.d.}(\tilde{G})$ and Theorem~\ref{thm:MAC} implies that
$(R_1-g_1,R_2-g_1, \dots ,R_n-g_1)\in C(\tilde{G})$ with $g_1=g_2+g_3$ which
gives the result in Theorem~\ref{thm:mainthm2}. A similar argument holds for
Theorem~\ref{thm:mainthm3}.

Note that the core of our argument in Propositions~\ref{prop0}, \ref{prop1},
and \ref{prop2} holds with no gap. The gaps in Theorems~\ref{thm:mainthm1},
\ref{thm:mainthm2} and \ref{thm:mainthm3} are due to the current approximation
gap of the capacity of Gaussian relay networks with respect to the i.i.d.
cutset upper bound.  Better approximations for the capacity of Gaussian relay
networks in terms of $C_{i.i.d.}$ can immediately improve the gap in our main
results. For example, in \cite{bobbie}, \cite{ritesh} and \cite{ritesh2} it is shown that
the approximations in \cite{AvDigTse09}, \cite{LKEC11}  can be significantly tightened for certain network configurations. 

Proposition~\ref{prop0} is proved in Section~\ref{sec:unicast} and Propositions~\ref{prop1} and \ref{prop2} are proved in Section~\ref{sec:MACandBC}.

\section{Unicast Networks}\label{sec:unicast}
In this section, we concentrate on proving Proposition~\ref{prop0}. We divide
our proof into two parts.  In the first part of the proof, we will show that
for any pair of links $(u,v)$ and $(v,u)$, we can remove one of the links
without changing $C_{i.i.d.}$.  Given this new network, we can iterate this
procedure for each bidirected link until we are left with a directed network
that contains no bidirected edges. In the second part of the proof we show that
given a directed network with cycles, we can remove at least one of the links
in the cycle without changing $C_{i.i.d.}$. Iterating this procedure for each
cycle, we can obtain a directed subnetwork  of the same $C_{i.i.d.}$ that
contains no cycles.

Our proof is based on two important properties of the cut function in \eqref{eq:cutfunction}:
\begin{enumerate}
\item For a fixed cut $S\subset V$, the cut values of a network $G$ and subnetwork $G'$ are
the same if all outgoing links from $S$ are in both $G$ and $G'$:
$$f(G;S) = f(G';S), \ \text{if } \forall (u,v) \in G:(u \in S, v \not \in S), (u,v) \in G'.$$
(Note that because $G'$ is a subgraph of $G$, the channel coefficients corresponding to $(u,v)$ are the same in both $G$ and $G^\prime$ if $(u,v)\in G^\prime$.)
\item $f(G;S)$ is a submodular function on $2^V$:
  $$f(G;S_1) + f(G;S_2) \ge f(G;S_1 \cup S_2) + f(G;S_1 \cap S_2) \qquad \forall S_1,S_2\subseteq V.$$
\end{enumerate}

The first property follows from the fact that when all outgoing links are in
both $G$ and $G'$, the MIMO matrix between $X_S$ and $Y_{\ol{S}}$ are the same,
and thus $I(X_S;Y_{\ol{S}} | X_{\ol{S}})$ which corresponds to the capacity of
this MIMO matrix is the same for both networks.  A proof of the second property
is given in Theorem 1 of \cite{submod}.

\subsection{Reduction of bidirected network to directed network}\label{directed}
\begin{figure}[t]
	\begin{center}
		\includegraphics[scale=0.45]{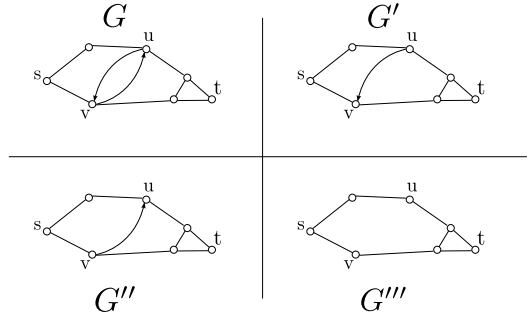}
	\end{center}
  \caption{Bidirected network with some of the links removed}
	\label{fig:1}
\end{figure}

Given a bidirected network $G$ and any pair of links $(u,v)$ and $(v,u)$, we
create the subnetworks $G'$, $G''$, and $G'''$, where the link $(v,u)$,
$(u,v)$, and both $(u,v)$ and $(v,u)$ are removed from $G$, respectively.  See
Figure \ref{fig:1}.

\begin{figure}[t]
	\begin{center}
		\includegraphics[scale=0.45]{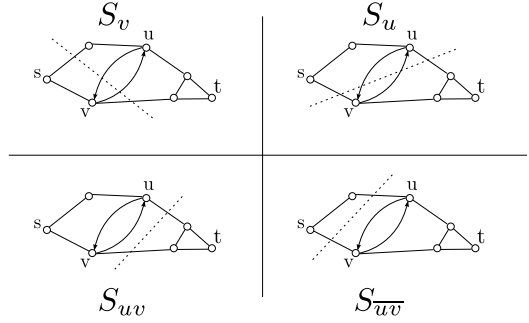}
	\end{center}
  \caption{Example of $S_v$, $S_u$, $S_{uv}$, $S_{\ol{uv}}$}
	\label{fig:3}
\end{figure}

Define $S_v$, $S_u$, $S_{uv}$, $S_{\ol{uv}}$, to be the following:
\begin{align*}
S_v &= \argmin_{\{S:s,v \in S\ t,u \not \in S\}} f(G;S)\\
S_u &= \argmin_{\{S:s,u \in S\ t,v \not \in S\}} f(G;S)\\
S_{uv} &= \argmin_{\{S:s,u,v \in S\ t \not \in S\}} f(G;S)\\
S_{\ol{uv}} &= \argmin_{\{S:s \in S\ t,u,v \not \in S\}} f(G;S).
\end{align*}
$S_v$ is the cut with the minimum cut value among all cuts for which $v$
remains on the source side and $u$ remains on the destination side; $S_u$ is
the cut with the minimum cut value among all cuts for which $u$ remains on the
source side and $v$ remains on the destination side; $S_{uv}$ is the cut with
the minimum cut value among all cuts for which both $u$ and $v$ are on the
source side; and $S_{\ol{uv}}$ is the cut with the minimum cut value among all
cuts for which $u$ and $v$ remain on the destination side. See
Figure~\ref{fig:3}. A cut that achieves the minimum cut value need not be
unique; we choose an arbitrary one in such cases.  Note that
\begin{equation}\label{eq:fourcuts}
C_{i.i.d.}(G)=\min \left(f(G;S_v),f(G;S_u), f(G;S_{uv}),f(G;S_{\ol{uv}})\right).
\end{equation}
We also define $S_v'$, $S_u'$, $S_{uv}'$, $S_{\ol{uv}}'$, and  $S_v''$,
$S_u''$, $S_{uv}''$, $S_{\ol{uv}}''$ in a similar fashion for graphs $G'$ and
$G''$, respectively.

Proposition~\ref{prop1} claims that either $C_{i.i.d.}(G)= C_{i.i.d.}(G')$ or
$C_{i.i.d.}(G)= C_{i.i.d.}(G'')$. We prove this by showing that each of the following assumptions lead to a contradiction:
\begin{itemize}
\item[(a)] $C_{i.i.d.}(G) < C_{i.i.d.}(G')$ and $C_{i.i.d.}(G) < C_{i.i.d.}(G'')$;
\item[(b)] $C_{i.i.d.}(G) < C_{i.i.d.}(G')$ and $C_{i.i.d.}(G) > C_{i.i.d.}(G'')$ (or  $C_{i.i.d.}(G) > C_{i.i.d.}(G')$ and $C_{i.i.d.}(G) < C_{i.i.d.}(G'')$);
\item[(c)] $C_{i.i.d.}(G) > C_{i.i.d.}(G')$ and $C_{i.i.d.}(G) > C_{i.i.d.}(G'')$.  
\end{itemize}
\medbreak 

\paragraph*{Case (a)} Assume $C_{i.i.d.}(G) < C_{i.i.d.}(G')$ and $C_{i.i.d.}(G) < C_{i.i.d.}(G'')$.

If $C_{i.i.d.}(G) < C_{i.i.d.}(G')$, then $C_{i.i.d.}(G)=f(G; S_v)$,
and 
\begin{equation}\label{eq:cont_a_1}
f(G; S_v)<\min \left(f(G;S_u), f(G;S_{uv}),f(G;S_{\ol{uv}})\right).
\end{equation}
This can be seen as follows. Note that the minimums in the definitions of $S_u,
S_{uv}$ and $S_{\ol{uv}}$ are taken over a set of cuts that cannot cross the
link $(v,u)$ and $G$ and $G^\prime$ only differ by the existence of the link
$(v,u)$. By Property (1), any cut that does not cross the edge $(v,u)$ has the
same value in $G$ and $G'$. Therefore, $f(G; S_u)=f(G^\prime; S_u)$, $f(G;
S_{uv})=f(G^\prime; S_{uv})$ and $f(G; S_{\ol{uv}})=f(G^\prime; S_{\ol{uv}})$.
Now, if the minimum in \eqref{eq:fourcuts} were to be achieved by any term
other than $f(G; S_v)$, this would imply that $C_{i.i.d.}(G^\prime)\leq
C_{i.i.d.}(G)$, which would contradict the assumption that $C_{i.i.d.}(G) <
C_{i.i.d.}(G')$. Therefore, we have \eqref{eq:cont_a_1}. 

Now, if also $C_{i.i.d.}(G) < C_{i.i.d.}(G'')$, by the same argument above we
should have  $C_{i.i.d.}(G)=f(G; S_u)$, and 
\begin{equation}\label{eq:cont_a_2}
f(G; S_u)<\min \left(f(G;S_v), f(G;S_{uv}),f(G;S_{\ol{uv}})\right).
\end{equation}
But \eqref{eq:cont_a_1} and \eqref{eq:cont_a_2} are contradictory.
\medbreak
\paragraph*{Case (b)} Assume $C_{i.i.d.}(G) < C_{i.i.d.}(G')$ and $C_{i.i.d.}(G) > C_{i.i.d.}(G'')$. Then, by the same argument in case (a), we have $C_{i.i.d.}(G)=f(G; S_v)$, and
\begin{equation}\label{eq:cont_b_1}
f(G; S_v)<\min \left(f(G;S_u), f(G;S_{uv}),f(G;S_{\ol{uv}})\right).
\end{equation}
Similarly, the assumption $C_{i.i.d.}(G) > C_{i.i.d.}(G'')$ implies that
$C_{i.i.d.}(G'')=f(G''; S_u'')$, and 
\begin{equation}\label{eq:cont_b_2}
f(G''; S_u'')<\min \left(f(G'';S_v''), f(G'';S_{uv}''), f(G'';S_{\ol{uv}}'')\right).
\end{equation}
This follows by the same argument for \eqref{eq:cont_a_1}: Since $G$ and $G''$
only differ by the existence of $(u,v)$, the value of the cut $S_u$ should be
different in $G$ and $G''$ while the values of the remaining three cuts are the
same in both $G$ and $G''$.

Note that \eqref{eq:cont_b_1} implies that
$$
f(G; S_v)<f(G;S_{\ol{uv}})\leq f(G; S_v\cap S_u''),
$$
where the last inequality follows from the fact that since $u\notin S_v$ and
$v\notin S_u''$,  $u,v\notin S_v\cap S_u''$ and the definition of $S_{\ol{uv}}$ which implies that among all such cuts of $G$, $S_{\ol{uv}}$ is the one with mincut value. Now, by Property (1), $f(G,
S_v)=f(G''; S_v)$ and $f(G; S_v\cap S_u'')=f(G''; S_v\cap S_u'')$ since $G$ and
$G''$ only differ by the existence of the link $(u,v)$ and both $S_v$ and
$S_v\cap S_u''$ correspond to cuts that cannot cross this link. Therefore, we
have $f(G''; S_v)< f(G''; S_v\cap S_u'')$.
On the other hand, \eqref{eq:cont_b_2} implies that
$$
f(G''; S_u'')<f(G'';S_{uv}'')\leq f(G''; S_v\cup S_u''),
$$
since $v\in S_v$ and $u\in S_u''$, $u,v\in S_v\cup S_u''$. Combining the last
two inequalities we obtain
$$
f(G''; S_v)+f(G''; S_u'')<f(G''; S_v\cap S_u'')+f(G''; S_v\cup S_u'').
$$
However, submodularity (Property (2)) for $f$ implies that 
$$
f(G''; S_v)+f(G''; S_u'')\geq f(G''; S_v\cap S_u'')+f(G''; S_v\cup S_u''),
$$
which leads to a contradiction.
\medbreak 

\paragraph*{Case (c)} Finally, we assume $C_{i.i.d.}(G) > C_{i.i.d.}(G')$ and $C_{i.i.d.}(G) > C_{i.i.d.}(G'')$. 

By similar arguments as in case (b), the first assumption implies that $C_{i.i.d.}(G') = f(G'; S_v')$, and the second one implies that  $C_{i.i.d.}(G'')=f(G''; S_u'')$. Moreover,
\begin{align}
f(G'; S_v') < f(G; S_v' \cup S_u'') \label{eq:6},
\end{align}
and  
\begin{align}
f(G''; S_u'') < f(G; S_v' \cap S_u''). \label{eq:7}
\end{align}
The last two inequalities follow from our assumption, $C_{i.i.d.}(G) >
C_{i.i.d.}(G')$ and $C_{i.i.d.}(G) > C_{i.i.d.}(G'')$, which implies that the
minimum cut values for $G'$ and $G''$ are strictly less than any cut value of
$G$. Combining (\ref{eq:6}) and (\ref{eq:7}), we have \begin{align*}
f(G'; S_v') + f(G''; S_u'') &< f(G; S_v' \cap S_u'') + f(G; S_v' \cup S_u''),
\end{align*}
Observing that by Property (1)
\begin{align}
f(G'; S_v') &=f(G'''; S_v') \\
f(G''; S_u'') &=f(G'''; S_u'')  \label{eq:2} \\
f(G; S_v' \cap S_u'') &= f(G'''; S_v' \cap S_u'') \label{eq:3} \\
 f(G; S_v' \cup S_u'') &=  f(G'''; S_v' \cup S_u'')\label{eq:4},
\end{align}
we  get
$$
f(G'''; S_v') + f(G'''; S_u'') < f(G'''; S_v' \cap S_u'') + f(G'''; S_v' \cup S_u'').
$$
This contradicts with the submodularity of $f$ in $G'''$.  Since cases (a), (b)
and (c) are eliminated, we conclude that either $C_{i.i.d.}(G)= C_{i.i.d.}(G')$
or $C_{i.i.d.}(G)= C_{i.i.d.}(G'')$.

\subsection{Removing cycles in a directed network}\label{cycle}
\begin{figure}[ht]
	\begin{center}
		\includegraphics[scale=0.55]{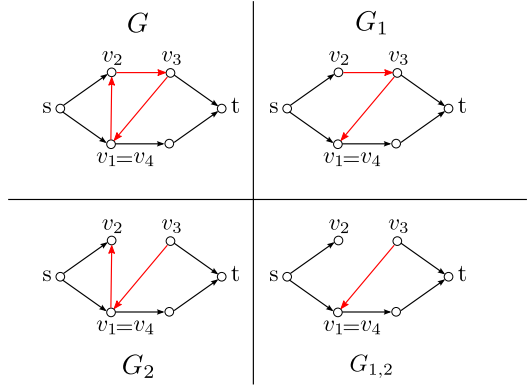}
	\end{center}
  \caption{An example of a directed network with a length 3 cycle and subnetworks with some links removed.}
	\label{fig:2}
\end{figure}

Consider a directed network $G$, where the nodes $\{v_1, v_2, \ldots v_N\}$
form a length $N$ cycle, and let $v_{N+1} = v_1$.  Define $G_k$, $k =
1,2,\ldots, N$ to be a subnetwork of $G$ with the link $(v_k,v_{k+1})$ removed.
In our proof, we denote subnetworks with both links $(v_{k-1},v_k)$ and
$(v_k,v_{k+1})$ removed as $G_{k-1,k}$.  See Figure (\ref{fig:2}) for an
example.  Let $S^*$ and $S_k$ denote cuts that achieve the minimum cut values
of networks $G$ and $G_k$, respectively:\looseness=-10
\begin{align*}
S^* &= \argmin_{\{S:s \in S\ t \not \in S\}} f(G;S), \\
S_k &= \argmin_{\{S:s \in S\ t \not \in S\}} f(G_k;S).
\end{align*}

We prove that $C_{i.i.d.}(G) = C_{i.i.d.}(G_k)$ for at least one value of $k$,
$k = 1,2,\ldots,N$ by showing that each of the following assumptions lead to a
contradiction:
\begin{itemize}
\item[(a)] $C_{i.i.d.}(G) > C_{i.i.d.}(G_k)$ for $k = 1,2,\ldots,N$;
\item[(b)] $C_{i.i.d.}(G) \not = C_{i.i.d.}(G_k)$ for $k = 1,2,\ldots,N$ and $C_{i.i.d.}(G) < C_{i.i.d.}(G_k)$ for at least one value of $k$.
\end{itemize}

\paragraph*{Case (a)} Assume $C_{i.i.d.}(G) > C_{i.i.d.}(G_k)$ for $k = 1,2,\ldots,N$.

Given our assumption, we first show that for each subnetwork $G_k$, there exists a cut
$S_k'$ that achieves the minimum cut value, i.e.,
\begin{equation}\label{eq:claim}
C_{i.i.d.}(G_{k})=f(G_k; S_{k}')
\end{equation}
with the property $v_1, v_2, \ldots, v_k \in
S_k'$ and $v_{k+1}\notin S_{k}'$.  This will lead to a contradiction when we take $k=N$.

If $C_{i.i.d.}(G) > C_{i.i.d.}(G_k)$, then $v_k \in S_k$ and $v_{k+1} \in
\ol{S_k}$.  This can be seen as follows.  Any cut that does not cross the link
$(v_k, v_{k+1})$ has the same cut value for both $G$ and $G_k$ by Property (1).
So the minimum cut value attained by $G_k$ must be for a cut which crosses the link
$(v_k, v_{k+1})$ and yields a cut value strictly less than any cut which does
not cross that link.  Thus, for $k=1$ we can choose $S_1' = S_1$. 

We will discover the sets $S_k'$ for larger $k$ by induction. We will show that
if the claim in \eqref{eq:claim} holds for $k-1$, it should also hold for $k$.

First note that since $S_{k-1}'$ and $S_k$ achieve the minimum cut values for
$G_{k-1}$ and $G_k$, they must be less than or equal to any other cut in
$G_{k-1}$ and $G_k$ respectively. In particular,
\begin{align}
\label{cycle:start1}
f(G_{k-1};S_{k-1}') &\le f(G_{k-1};S_{k-1}' \cap S_k), \\
f(G_k;S_k) &\le f(G_{k};S_{k-1}' \cup S_k).
\label{cycle:end1}
\end{align}
Now, since $v_k \in
\ol{S_{k-1}'}$ and $v_k \in \ol{S_{k-1}' \cap S_k}$, $(v_k,v_{k+1})$ cannot be
an outgoing link in either of the cuts $S_{k-1}'$ and $S_{k-1}'\cap S_k$, and
all other links in $G_{k-1}$ are also in $G_{k-1,k}$, so by Property (1) of
$f$, we have
\begin{align} \label{cycle:start2}
f(G_{k-1,k};S_{k-1}') &= f(G_{k-1};S_{k-1}'), \\
f(G_{k-1,k};S_{k-1}' \cap S_k) &= f(G_{k-1};S_{k-1}' \cap S_k).
\end{align}
Also, $v_k \in S_k$ and $v_k \in S_{k-1}' \cup S_k$, so $(v_{k-1},v_k)$ cannot
be an outgoing link in either of those cuts, and all other links in $G_k$ are
also in $G_{k-1,k}$.  So again by Property (1) of $f$, we have\looseness=-1
\begin{align}
f(G_{k-1,k};S_k) &= f(G_k;S_k),\\
f(G_{k-1,k};S_{k-1}' \cup S_k) &= f(G_k;S_{k-1}' \cup S_k).
\label{cycle:end2}
\end{align}
By the submodular property of $f$ on $G_{k-1,k}$ we have 
\begin{align*}
f(G_{k-1,k}; S_{k-1}') &+ f(G_{k-1,k}; S_k) \ge\\
f(G_{k-1,k}; S_{k-1}' \cap S_k) &+ f(G_{k-1,k}; S_{k-1}' \cup S_k),
\end{align*}
and equations
(\ref{cycle:start2})-(\ref{cycle:end2}) yield
\begin{align*}
f(G_{k-1}; S_{k-1}') &+ f(G_k; S_k) \ge\\
f(G_{k-1}; S_{k-1}' \cap S_k) &+ f(G_k; S_{k-1}' \cup S_k).
\end{align*}
Combining this result and equations (\ref{cycle:start1}) and (\ref{cycle:end1}) yields
\begin{align*}
f(G_k; S_k) = f(G_k; S_{k-1}' \cup S_k).
\end{align*}
Thus the cut $S_k' = S_{k-1}' \cup S_k$ achieves the minimum cut value for
network $G_k$ and has the property $v_1, \ldots, v_k \in S_k'$.  Now suppose
$v_{k+1} \in S_k'$.  Then the cut $S_k'$ cannot cross the link $(v_k,
v_{k+1})$, and thus $f(G_k; S_k') = f(G; S_k')$.  But since $S_k'$ achieves the
minimum cut value for network $G_k$, we have the following:
\begin{align*}
C_{i.i.d}(G_k) = f(G_k; S_k') = f(G; S_k') \ge C_{i.i.d}(G),
\end{align*}
which would contradict our assumption $C_{i.i.d}(G)>C_{i.i.d}(G_k)$.  The last inequality follows because
$C_{i.i.d}(G)$ must be less than or equal to any cut value of $G$.  Thus,
$v_{k+1} \not \in S_k'$.  

Letting $k=N$, we have $v_1, \ldots, v_N \in S_N'$, but $v_{k+1}=v_1 \not \in
S_N'$, which is a contradiction.
\bigbreak

\paragraph*{Case (b)} Assume $C_{i.i.d.}(G) \not = C_{i.i.d.}(G_k)$ for $k = 1,2,\ldots,N$ and $C_{i.i.d.}(G) < C_{i.i.d.}(G_k)$ for at least one value of $k$.

Without loss of generality, let $C_{i.i.d.}(G) < C_{i.i.d.}(G_1)$.
By the same arguments as in the previous case, $v_1 \in S^*$, $v_2 \in
\ol{S^*}$.

We now show that for $k > 1$, if $v_k \in \ol{S^*}$, then $v_{k+1}
\in \ol{S^*}$. This will lead to a contradiction when we take $k=N$.

Assume $v_k \in \ol{S^*}$ and consider $C_{i.i.d}(G_k)$:
\begin{align*}
C_{i.i.d}(G) &= f(G; S^*) \\
&\overset{(a)}{=} f(G_k; S^*)  \\
&\overset{(b)}{>} C_{i.i.d}(G_k).
\end{align*}
(a) follows by Property (1) and the fact that $v_k \in \ol{S^*}$, and so $S^*$
cannot cross the link $(v_k, v_{k+1})$.  (b) follows from the fact that
$C_{i.i.d}(G_k)$ must be less than or equal to any cut value of $G_k$, i.e. $f(G_k; S^*)\geq C_{i.i.d}(G_k)$  and our
assumption that $C_{i.i.d}(G) \ne C_{i.i.d}(G_k)$ for $k = 1,2,\ldots,N$, thus
making the inequality strict. Now since $C_{i.i.d}(G_k)<C_{i.i.d}(G)$, for the
minimum cut $S_k$, we must have $v_k \in S_k$, $v_{k+1} \in \ol{S_k}$.  Next,
consider $S^* \cap S_k$.  We have
\begin{align}
\label{cycle:start3}
f(G_k; S^* \cap S_k) &\overset{(a)}{=} f(G; S^* \cap S_k)\\
& \overset{(b)}{>} f(G_k;S_k),
\end{align}
where (a) follows by Property (1) and the fact that $v_k \in \ol{S^* \cap
S_k}$, and so $S^*\cap S_k$ cannot cross the link $(v_k, v_{k+1})$.  (b)
follows because $C_{i.i.d}(G_k) < C_{i.i.d}(G)$, so the minimum cut value of
$C_{i.i.d}(G_k) = f(G_k;S_k)$, must be strictly less than any cut value of $G$.
Now suppose $v_{k+1} \in S^*$.  Then
\begin{align}
f(G_k; S^* \cup S_k) &\overset{(a)}{=} f(G; S^* \cup S_k)\\
& \overset{(b)}{\ge} f(G;S^*) \\
& \overset{(c)}{=} f(G_k;S^*).
\label{cycle:end3}
\end{align}
(a) and (c) follow because $v_{k+1} \in S^* \cup S_k$ and  $v_{k+1} \in S^*$,
so neither of those cuts can cross the link $(v_k, v_{k+1})$.  (b) follows
because $S^*$ achieves the minimum cut value of graph $G$.

Combining (\ref{cycle:start3})-(\ref{cycle:end3}), we have
\begin{align*}
f(G_k; S^* \cup S_k) + f(G_k; S^* \cap S_k) > f(G_k;S_k) + f(G_k;S^*).
\end{align*}
This contradicts the submodularity of $f$ in $G_k$.  Thus $v_{k+1} \in
\ol{S^*}$.  

Letting $k = N$, we have $v_1 \in S^*, v_2, v_3, \ldots, v_N \in \ol{S^*}$.
However, because $v_1,\dots, v_N$ form a cycle, the node $v_1$ can be thought
of $v_{N+1}$ and $v_N \in \ol{S^*}$ by the above iteration implies that $v_1
\in \ol{S^*}$. This contradicts with the fact that $v_1 \in S^*$ and shows that
the initial assumptions for case (b) necessarily lead to a contradiction.

Since we have eliminated cases (a) and (b) above, we conclude that
$C_{i.i.d}(G) = C_{i.i.d}(G_k)$ for at least one value of $k = 1,2,\ldots,N$.

\section{Multiple Access and Broadcast Networks}\label{sec:MACandBC}
\subsection{Multiple Access Networks}
In this section, we use Proposition~\ref{prop0} to prove
Proposition~\ref{prop1}. Consider a Gaussian network $G$ with multiple access
traffic between the sources $s_1,s_2,\ldots s_n$ and the destination $d$.
Assume that $(R_1,R_2,\ldots R_n) \in C_{i.i.d.}(G)$. We will show that there exists an acyclic subnetwork $\tilde{G}$ of $G$ such that $(R_1,R_2,\ldots R_n) \in C_{i.i.d.}(\tilde{G})$.

Starting from $G$, we first construct an extended directed graph $G^{'}=(V^{'},E^{'})$ as follows:
\begin{enumerate}
\item Let $V^{'}=V \cup \lbrace s^{'} \rbrace$ where $s^{'}$ is an added auxiliary vertex.
\item Let $E^{'}=E\cup \lbrace (s^{'},s_i) \mid 1\leq i\leq n \rbrace$.
\end{enumerate}
We assume that each edge $(s^{'},s_i)$ represents an isolated edge of capacity
$R_i$, for $i=1,2,\dots,n$. This can be done within the Gaussian network model
we defined in Section~\ref{sec:model} by assuming, for example, that $s^{'}$ is
equipped with $n$ transmit antennas where  each transmit antenna is connected
only to the corresponding $s_i$ with a Gaussian channel of capacity $R_i$ (the
channel coefficient of this channel is chosen accordingly). Consider this
Gaussian network $G^{'}$ with unicast traffic from $s^{'}$ to $d$. We next
lower bound $C_{i.i.d.}(G^{'})$ for this unicast network. Note that for any cut $S\subseteq V^{'}:
s^{'}\in S,$
\begin{equation}
\begin{split}
f(G^{'};S)&=f(G;S\setminus \lbrace s^{'}\rbrace )+\sum_{s_i\notin S}{R_i}\\ 
&\geq \sum_{s_i\in S}{R_i}+\sum_{s_i\notin S}{R_i}\\ 
&=\sum_{i=1}^{n}{R_i},
\end{split}
\end{equation}
where the first equality follows from the fact that $f(G^{'};S)=I(X_{S};Y_{S^c}\lvert
X_{S^c})$ under i.i.d. input distributions decomposes into $f(G^{'};S\setminus \lbrace s^{'}\rbrace )+\sum_{s_i\notin S}{R_i}$ since $(s^{'},s_i)$ are isolated from other channels in $G^{'}$ and also from each other. In turn, $f(G^{'};S\setminus \lbrace s^{'}\rbrace )=f(G;S\setminus \lbrace s^{'}\rbrace )$ since due to the way we constructed $G^{'}$ the outgoing edges from $S\setminus \lbrace s^{'}$ are the same in both $G$ and $G^{'}$.  The second line follows from our assumption that $(R_1,R_2,\ldots R_n) \in C_{i.i.d.}(G)$ and the definition of $C_{i.i.d.}(G)$ for multiple access networks in \eqref{eq:cutfunction1}. Therefore, we can conclude
that for the constructed unicast network $G^{'}$, we have $$C_{i.i.d.}(G^{'})\geq \sum_{i=1}^{n}{R_i} .$$

Now, due to Proposition~\ref{prop0}, we know that can find an acyclic subnetwork
${\tilde{G}}^{'}$ of $G^{'}$ for which
$C_{i.i.d.}({\tilde{G}}^{'})=C_{i.i.d.}(G^{'})$. Let ${\tilde{G}}$ be the graph
obtained by removing $s^{'}$ and $\lbrace (s^{'},s_i) \mid 1\leq i\leq n \rbrace$ from ${\tilde{G}}^{'}$. Note that ${\tilde{G}}$ is an acyclic subnetwork of our original multiple access network $G$. To complete the proof, we
show that $(R_1,R_2, \dots ,R_n)\in
C_{i.i.d.}({\tilde{G}})$. Consider an arbitrary $S \subseteq V$. Let $S^{'}=S
\cup \lbrace s^{'} \rbrace$. Since $C_{i.i.d.}({\tilde{G}}^{'})$ is not less
than $\sum_{i=1}^{n}{R_i}$, we have
\begin{equation}
\begin{split}
f({\tilde{G}};S) &= f({\tilde{G}};S)+\sum_{s_i \notin S}{R_i} -\sum_{s_i \notin S}{R_i}
\\ &= f({\tilde{G}}^{'};S^{'})-\sum_{s_i \notin S}{R_i}
\\ &\geq \sum_{i=1}^{n}{R_i} - \sum_{s_i \notin S}{R_i}
\\ &= \sum_{s_i \in S}{R_i}.
\end{split}
\end{equation}
where the second equality again follows from the fact that $f({\tilde{G}}^{'};S^{'})$ decomposes into $f({\tilde{G}}^{'};S)+\sum_{s_i \notin S}{R_i}$ and $f({\tilde{G}};S)=f({\tilde{G}}^{'};S)$. Thus, according to the definition of $C_{i.i.d.}({\tilde{G}})$ for a multiple access network in
\eqref{eq:cutfunction1} we have shown that $(R_1,R_2, \dots ,R_n)\in
C_{i.i.d.}({\tilde{G}})$, and this completes the proof for
Proposition~\ref{prop1}.

\begin{figure}[t]
	\begin{center}
		\includegraphics[scale=0.5]{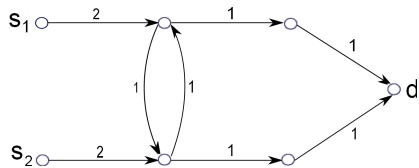}
	\end{center}
  \caption{An example of a multiple-access network where no acyclic subnetwork achieves the capacity region of the original network. Observe that while both rate points $(2,0)$ and $(0,2)$ are in capacity region of the original network, there is no acyclic subnetwork that achieves both of these rate points.}
	\label{fig:MACE}
\end{figure}

Note that although we have proved that every rate tuple in the capacity region of a multiple access network can be achieved by using an acyclic subnetwork, we cannot conclude that there exists an acyclic subnetwork which has the same capacity region as the
original network. For an example, consider the network in
Figure~\ref{fig:MACE} which depicts a multiple access network with isolated edges of
corresponding capacities. Observe that both $(R_1, R_2)=(2,0)$ and $(R_1, R_2)=(0,2)$ are in the capacity region of the original network, however neither of two acyclic
subnetworks can have both of these rate points in its capacity region. In other words,
despite the fact that for each achievable rate point there exists an acyclic
subnetwork achieving that rate point, these subnetworks may differ for
different rate points, leading to cases where the capacity regions of all
the acyclic subnetworks of a network are strictly smaller than the capacity of the original
network.   

\subsection{Broadcast Networks}
\begin{figure}[t]
	\begin{center}
		\includegraphics[scale=0.5]{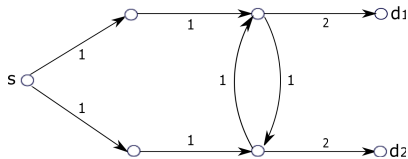}
	\end{center}
  \caption{An example of a broadcast network where no acyclic subnetwork achieves the capacity region of the original network. Observe that while both rate points $(2,0)$ and $(0,2)$ are included in capacity region of the original network, there is no acyclic subnetwork that achieves both of these rate points.}
	\label{fig:MABC}
\end{figure}
Proposition~\ref{prop2} for broadcast traffic can be proved by using a similar
approach to Proposition~\ref{prop1}.  Consider a Gaussian network $G=(V,E)$ with
broadcast traffic where source $s$ communicates independent messages to
destinations $d_1,d_2,\dots ,d_n$. Let $C_{i.i.d.}(G)$ be the associated rate
region and let $(R_1,R_2,\dots R_n)\in C_{i.i.d.}(G) $. As before, we first create
a unicast network $G^{'}=(V^{'},E^{'})$ from $G=(V,E)$ by adding an auxiliary vertex
$d^{'}$ to $G$ such that $V^{'}=V \cup \lbrace d^{'}\rbrace$, $E'=E\cup \lbrace
(d_i,d') \vert 1\leq i\leq n\rbrace$. Let each $(d_i,d')$ be an isolated edge
of capacity $R_i$. Then for any $s-d'$ cut of the unicast network $G'$,  $S \subseteq
V'$ and $d'\notin S$, we have 
\begin{equation}
\begin{split}
f(G';S)&=f(G;S)+\sum_{d_i \in S}{R_i}
\\ &\geq \sum_{d_i \notin S}{R_i} + \sum_{d_i \in S}{R_i} = \sum_{i=1}^{n}{R_i},
\end{split}
\end{equation}
where the inequality follows from our assumption that $(R_1,R_2,\dots R_n)\in C_{i.i.d.}(G) $. Therefore, $C_{i.i.d.}(G')\geq \sum_{i=1}^{n}{R_i} $ and using
Proposition~\ref{prop0} we can find an acyclic subnetwork ${\tilde{G}}^{'}$ in
$G'$ for which $C_{i.i.d.}({\tilde{G}}^{'})= C_{i.i.d.}(G')\geq
\sum_{i=1}^{n}{R_i}$. Let ${\tilde{G}}$ be the broadcast network obtained by removing
the additional node $d^{'}$ and the edges $(d_i,d')$ from ${\tilde{G}}^{'}$. As before, we can argue that $(R_1,R_2,\dots
R_n)\in C_{i.i.d.}(\tilde{G})$. For any $S\subseteq V$, we have
\begin{equation*}
\begin{split}
f(\tilde{G};S)&=f(\tilde{G};S)+\sum_{d_i \in S}{R_i}-\sum_{d_i \in S}{R_i}
\\ &= f(\tilde{G}^{'};S) - \sum_{d_i \in S}{R_i}
\\ &\geq \sum_{d_i \notin S}{R_i}.
\end{split}
\end{equation*}
Thus, $(R_1,R_2,\dots R_n)\in C_{i.i.d.}(\tilde{G}) $, and the proof of
Proposition~\ref{prop2} is complete.

Note that again as in the case of multiple access traffic, the above result does not imply the existence of a single acyclic subnetwork whose capacity region is as large as
the original network.  For a counter example one can consider the network in
Figure~\ref{fig:MACE} with the directions of the edges reversed. See
Figure~\ref{fig:MABC}.

\section{Multicast and Multiple Unicast Networks}
As opposed to the multiple access and broadcast networks discussed in the
earlier sections, bidirected communication across certain links can be
necessary to achieve certain rate points in the capacity regions of multicast
and multiple unicast networks. For multicast, consider the network in
Figure~\ref{fig:MABC}, but assume that the source wants to multicast the same
information to both of the destination nodes. The multicast capacity of this
network is $2$, however the multicast capacity of any of its acyclic
subnetworks is equal to $1$. For the multiple unicast case, the classical
Gaussian interference channel with feedback readily provides an example where
feedback (i.e. bidirected communication) is necessary for achieving capacity.
It also straightforward to construct simple examples of multiple unicast
networks with isolated edges which illustrate this point.

\section{Concluding Discussion}\label{sec:discuss}
\begin{figure}[t]
	\begin{center}
		\includegraphics[scale=0.5]{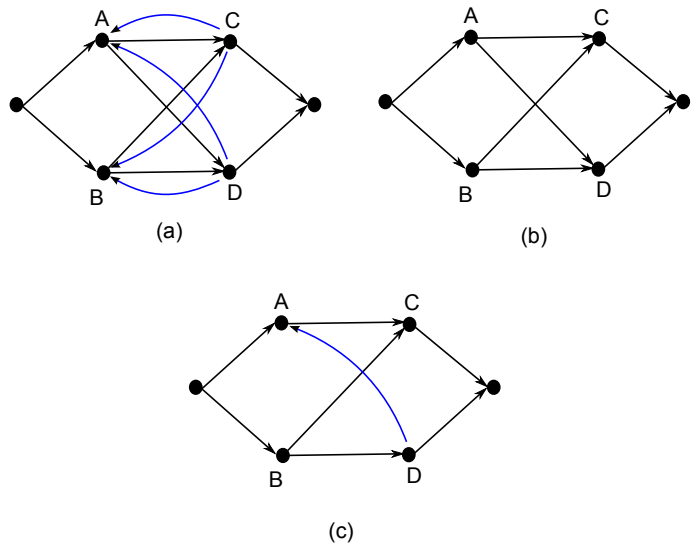}
	\end{center}
  \caption{Bidirected network with some of the links removed}
	\label{fig:twolayers}
\end{figure}

In this paper, we discussed the usefulness of feeding back information through cycles in Gaussian multi-hop networks. We showed that for unicast, broadcast and multiple-access networks, every rate point in the capacity region of the original network can be approximately achieved in a cycle-free manner, i.e. by using an acyclic subnetwork of the original network. The approximation here is within a bounded gap which is independent of the channel coefficients and the SNRs in the network which implies that feeding back information through cycles in such networks can only provide a bounded improvement in achievable rates as SNR grows. 

As studied in \cite{ACLY00} and \cite{LKEC11}, cycles significantly increase
the delay and complexity of (approximately) optimal relaying strategies. By
identifying a directed acyclic subnetwork that is sufficient to approximately
maintain capacity, our result can be used to reduce the delay and complexity of
such schemes by suggesting links that could be potentially shut down.
Although shutting down individual links in wireless networks may be
nontrivial since these links may represent overheard transmissions over other
links, certain networks such as Gaussian networks consisting of isolated links or only MAC and broadcast components (as studied in \cite{KoetterEffrosMedard} and
\cite{KanVis11}) provide some freedom in controlling individual links. Indeed,
simplification can be possible even in more general networks.

Consider the example in Figure~\ref{fig:twolayers}-(a) where the edges in the
graph indicate the wireless links with non-zero channel gains. Assume that the
backward links from the second layer of relays (nodes C and D) to the first
(nodes A and B) operate over a separate frequency, so that while signals
arriving over the same colored edges superpose at a node, signals over
different colored edges arrive separately.  Similarly, while signals over the
same colored edges emanating from a single node represent broadcast, different
signals can be transmitted over different colored edges. If the directed
acyclic network in Figure~\ref{fig:twolayers}-(b) is identified as sufficient for
preserving the capacity of the network, this implies that the backward channel
from the second layer to the first need not be used at all. On the other hand,
if the directed subnetwork is the one in (c), there is no operational way to
reduce the wireless network in (a) to (c).  The forward link from node A to D
cannot be avoided. However, the communication over the backward channel can
still be simplified by not transmitting over the blue frequency from node C and by ignoring the received signal over the blue frequency at node B.


\begin{thebibliography}{99}
\bibitem{Shan56} C. E. Shannon, \emph{The zero error capacity of a noisy
  channel}, IRE Transactions on Information Theory, September 1956.

\bibitem{EGC79} T.~M.~Cover and A.~El~Gamal, \emph{Capacity theorems for the
  relay channel}, IEEE Trans. on Information Theory, vol. 25, no. 5,
  pp.572--584, September 1979.

  
\bibitem{Ozarow84} L. H. Ozarow, \emph{The capacity of the white Gaussian
  multiple access channel with feedback}, IEEE Trans. on Information Theory,
  vol.  30, no. 4, pp.623--629, July 1984.

\bibitem{Thomas87} J.~Thomas, \emph{Feedback can at most double Gaussian multiple access channel capacity}, IEEE Trans. on Information Theory, vol. 33, no. 5,
  pp.711 --716, September 1987.
  
\bibitem{SuhTse11} C. Suh and D. Tse, \emph{Feedback capacity of the Gaussian
  Interference channel to within 2 bits}, IEEE Trans. on Information Theory,
  vol.  57, no. 5, pp 2667--2685, May 2011.

\bibitem{OLT07} A.~{\"O}zg{\"u}r, O.~L{\'e}v{\^e}que, D.~Tse,
  \emph{Hierarchical Cooperation Achieves Optimal Capacity Scaling in Ad-Hoc
  Networks}, IEEE Trans. on Information Theory, vol. 10, no. 53, pp.3549--3572,
  October 2007.

\bibitem{AvDigTse09} S.~Avestimehr, S N.~Diggavi and D.~Tse, \emph{Wireless
  network information flow: a deterministic approach}, IEEE Trans. on
  Information Theory, vol. 57, no. 4, pp 1872--1905, April 2011.  

\bibitem{KoetterEffrosMedard} R. Koetter, M. Effros, and M. M\'{e}dard,
  \emph{A theory of network equivalence - Part II: Multiterminal Channels},
  Available online at arXiv:1007.1033.

\bibitem{simplification} C. Nazaroglu, A. \"{O}zg\"{u}r, and C. Fragouli,
  \emph{Wireless Network Simplification: the Gaussian N-Relay Diamond Network},
  IEEE Int. Symposium on Information Theory (ISIT), St Petersburg, 2011. 

\bibitem{michelle} S. Jalali, M. Effros, and T. Ho, \emph{On the impact of a
  single edge on the network coding capacity}, Information Theory and
  Applications Workshop (ITA), 2011. 

\bibitem{OD10} A.~Ozgur and S~N. Diggavi, \emph{Approximately achieving
  Gaussian relay network capacity with lattice codes}, IEEE Int. Symposium on
  Information Theory (ISIT), Austin, 2010.

\bibitem{submod} F. Parvaresh and R. Etkin,  \emph{Efficient Capacity
  Computation and Power Optimization for Relay Networks}, Submitted to IEEE
  Trans. Info. Theory, 2011.

\bibitem{bobbie} B. Chern and A. \"{O}zg\"{u}r, \emph{Achieving the capacity of
  the $N$-relay Gaussian diamond network within $\log N $ bits}, IEEE
  Information Theory Workshop, Lausanne, 2012.

\bibitem{ritesh} R. Kolte and A. \"{O}zg\"{u}r, \emph{Improved capacity
  approximations for Gaussian relay networks}, IEEE Information Theory
  Workshop, Seville, 2013.

\bibitem{ritesh2} R. Kolte, A. \"{O}zg\"{u}r and A. El Gamal, \emph{Optimized Noisy Network Coding for Gaussian Relay Networks}, IEEE International Zurich Seminar on Communications, 2014.

\bibitem{ACLY00} R. Ahlswede, N. Cai, S.-Y. R. Li, R. W. Yeung, \emph{Network
  Information Flow}, IEEE Trans. on Information Theory, vol.~46, no.~4, p.
  1204-1216, July 2000.

\bibitem{LKEC11} S.~Lim, Y.-H. Kim, A.~El-Gamal, and S-Y.Chung, \emph{Noisy
  network coding}, IEEE Trans. on Information Theory, vol.~57, no.~5, p.
  3132--3152, May 2011.

\bibitem{KanVis11} S. Kannan, A. Raja and P. Viswanath,  \emph{Local Phy +
  Global Flow: A Layering Principle for Wireless Networks}, IEEE Int. Symposium
  on Information Theory (ISIT), St Petersburg, 2011; e-print
  http://arxiv.org/abs/1111.4768.

\bibitem{broadcast} S. Kannan, A. Raja and P. Viswanath,  \emph{Approximately
Optimal Wireless  Broadcasting}, IEEE Trans. on Information Theory, vol.~58,
no.~12, p.  7154--7167, DEc. 2012.


\end{thebibliography}
\end{document}